 \documentclass[12pt, draftclsnofoot, onecolumn]{IEEEtran}
 \normalsize
\usepackage{epsfig}
\usepackage{amsmath}
\usepackage{theorem}
\usepackage{amsmath}
\usepackage{mathrsfs}
\usepackage{epsfig}
\usepackage{amsfonts}
\usepackage{amssymb}
\usepackage{tabularx}
\usepackage{mathrsfs}
\usepackage{euscript}
\usepackage{graphicx}
\usepackage{multirow}
\usepackage{stfloats}
\usepackage{cite}

\usepackage{color}

\usepackage{subcaption}

\begin{document}
	\title{Spectral Analysis of GFDM Modulated Signal under Nonlinear Behavior of Power Amplifier }
	\author{Amirhossein Mohammadian$^1$, Abbas Mohammadi$^1$, Senior Member, IEEE, and Abdolali Abdipour$^1$, Senior Member, IEEE, Mina Baghani$^1$ 
		\thanks{$^{(1)}$ Microwave / Millimeter-Wave and Wireless Communications Research Lab., Electrical Engineering Department, Amirkabir University of Technology, Tehran, Iran (e-mail: {amirmohammadian, abm125, abdipour, baghani}@aut.ac.ir).}
}
\markboth{}%
{}	
	\maketitle	
	\begin{abstract}
		General frequency division multiplexing (GFDM) is a flexible non-orthogonal waveform candidate for 5G which can offer some advantages such as low out-of-band (OOB) emission and high spectral efficiency. In this paper, the effects of nonlinear behavior of practical PAs on GFDM signal are studied. In the first step, a closed form expression for power spectral density (PSD) of GFDM signal is extracted. Then, the PSD at the output of PA as a function of input power and the coefficients of nonlinear polynomial PA model is derived. In addition, the adjacent channel power (ACP) and ACP ratio, as two important performance metrics, are evaluated. The simulation results confirm the accuracy of derived analytical expressions. Moreover, to validate the performance of GFDM modulation after nonlinear PA, it is compared with OFDM modulation.
	\end{abstract}
	
	\begin{keywords}
		5G, GFDM modulation, Power amplifier nonlinearity, Power spectral density, Adjacent channel interference, Saturation and $1\textrm{dB}$ compression points
	\end{keywords}

	\IEEEpeerreviewmaketitle
	
	\section{Introduction}

	 \IEEEPARstart{W}{ide} range of applications such as internet of things (IoT), machine-to-machine (M2M) communication, vehicle-to-vehicle (V2V) communication, tactile internet and cognitive radio are turning points in emergence of next generation wireless cellular networks (5G)\cite{magh8}. For accomplishing these applications implementation, the next generation of the wireless systems must overcome the challenges such as lower power consumption, higher spectral efficiency, lower out-of-band emission and lower latency compared to the previous generation of wireless communication systems \cite{magh8,magh9}. Indispensable role of modulation technique in addressing these requirements is undeniable.
	
Orthogonal frequency division multiplexing (OFDM) scheme is a common multi-carrier modulation which is used in physical layer of the modern cellular communication system \cite{magh2}. In spite of its simple implementation\cite{magh11}, it cannot deal with all vital requirements of 5G. Solid synchronization requirement in order to keep the orthogonality of OFDM subcarriers impose higher power consumption\cite{magh8}. Moreover, by adding cyclic prefix to each OFDM symbol, obtaining high throughput is not easily achievable\cite{magh12}. In addition, high out-of-band (OOB) emission of OFDM\cite{magh13} is absolutely inefficient for cognitive radio networks and multiple access scenarios. Thus, the limitations of OFDM modulation in satisfying these requirements forces next generation networks to use a new modulation technique\cite{magh8}. Filter bank multicarrier (FBMC)\cite{magh14} and generalized frequency division multiplexing (GFDM)\cite{magh15} are two potential candidates of non-orthogonal waveform for 5G networks. Even though, FBMC has ultra low OOB and high spectral efficiency, it has some drawbacks including difficulty of equalization process for frequency selective fading channel and its complexity for MIMO implementation\cite{magh47,magh48}. Hence, we consider GFDM waveform in this paper. 
	 
	 GFDM is a flexible non-orthogonal multi-carrier modulation which is proposed in\cite{magh15}. As a block-based modulation scheme, GFDM includes number of subcarriers, each one carries subsymbols generated in multiple time slots. The subcarriers are individually pulse-shaped with a prototype filter using circular convolution. Flexibility of GFDM pertains to degrees of freedom measured by prototype filter and number of subcarriers and subsymbols. As a consequence of selecting pulse shaping filter, OOB emissions of GFDM degrades and therefore makes it attractive for noncontiguous frequency bands. In addition, the synchronization requirements of GFDM could be relaxed by utilizing additional cyclic suffix \cite{magh15}. Moreover, by adding cyclic prefix to the entire block instead of each timeslot, higher spectral efficiency is achieved in compare with OFDM modulation.

	GFDM modulation as a multicarrier modulation based on filter bank concept was initially discussed in\cite{magh16}. The matrix model of GFDM transmitter was derived in linear form in\cite{magh17}. 
	 Moreover, by using different prototype filters, the BER and OOB emission of GFDM were studied and the effectiveness of pulse shaping on GFDM performance was studied in\cite{magh18}. The adaptation of GFDM to 5G applications was shown in \cite{magh15} by analyzing many characteristics of it. 
	Although GFDM is a promising candidate for 5G, its practical implementation is confined with some impairments. One of them is high complexity compared to OFDM which is a major disadvantage. Accordingly, the complexity in the receiver was reduced by representing GFDM in frequency domain and employing sparse feature of filter in \cite{magh23}. Also, a modem structure with low computational cost based on fast Fourier transform (FFT) was recently presented in \cite{magh24}. Nonlinear nature of power amplifier (PA) is another practical impairment which may influence the GFDM performance.
	
	PA is one of the most important elements of radio frequency (RF) chain in all wireless communication systems\cite{magh26}. The PA linearity and efficiency are two conflicting but important factors. In other words, by moving the operating point towards the saturation point, PA efficiency increases while its linearity decreases\cite{magh27}. According to increase in power consumption of communication systems in which PA has a huge part, utilizing the PA in high efficiency region is necessitated. Therefore, investigating the influence of PA's nonlinear behavior on system's performances is pivotal. Due to the nonlinear behavior of PA, AM/AM and AM/PM distortions may occur\cite{magh28}. These destructive effects should be investigated for different types of input signal including GFDM. The authors in\cite{magh25} has designed a receiver which estimates noise generated by in-band distortion of nonlinear PA and compensates its effects. In addition to in-band distortion, nonlinear nature of PA causes out-of-band distortion but to the best of our knowledge, none of the articles has investigated the GFDM out-of-band spectrum expansion, called spectral regrowth \cite{magh29}. This issue is critical in multiple access scenarios and specially cognitive radio networks in which unlicensed users are not allowed to produce interference on the active licensed users on the adjacent channels\cite{magh30}. This motivates us to investigate the power leakage of GFDM on the adjacent channels due to the nonlinear PA.

	Many articles have focused on the effects of nonlinear behavior of PA for both single carrier and multicarrier input signals. The spectral regrowth of stationary complex Gaussian input signal was derived by calculating the autocorrelation function of PA output signal\cite{magh31,magh32,magh33,magh34,magh35}. Indeed, code devision multiple access (CDMA) in \cite{magh31,magh34}, and OFDM in \cite{magh32,magh35} satisfied the conditions of input complex Gaussian distribution according to the central limit theorem for high number of users and subcarriers, respectively. Without considering Gaussian distribution for PA input signal, the autocovariance function of PA output was derived in\cite{magh36}. In fact, assuming PA input to be a stationary process is not valid in most cases. Accordingly, the PSD expression of PA output was precisely derived for cyclostationary signals in\cite{magh37}. More generally, by assuming orthogonality between subcarriers and time-limited pulse shape, a mathematical model of PSD for multi-carrier signals was derived in\cite{magh38} which by considering its assumptions, the method can not be applied to GFDM. Finally, these analysis should be considered in system design especially in cognitive radio network. Thus, the results of nonlinearity analysis in single carrier and multicarrier were used in \cite{magh39} and\cite{magh40} for resource allocation in cognitive radio network, respectively.

	In this paper, the PA nonlinearity distortion on the GFDM modulated signal is analyzed. For this purpose, the closed-form expressions for PSD of GFDM signal is derived. In the following, we assume that the nonlinear behavior of PA is modeled by polynomial function. Then, the PSD of nonlinear PA output signal is analyzed. Accordingly, closed-form expression of two critical points known as $1\textrm{dB}$ compression and saturation points are calculated. Moreover, the adjacent channel power (ACP) and the adjacent channel power ratio (ACPR) metrics which show basic information about power leakage in the adjacent channel are considered. Finally, to indicate the performance of GFDM after passing through nonlinear PA, it is compared with OFDM.
	
	The rest of this paper is organized as follows: a system model 
	is presented in Section  \ref{sec:model}. The PSD and autocorrelation function of GFDM signal and PA output are calculated in Section \ref{sec:bachground}. The ACP and ACPR, $1\textrm{dB}$ compression and saturation points are derived in Section \ref{sec:1db}. The accuracy of  analytical expressions are examined in Section \ref{sec:sim} by means of simulation results. Finally, conclusion remarks are provided in Section \ref{sec:conclusion}.
	
	\begin{figure*}[t!]
		\begin{center}
			\includegraphics[width=1\textwidth]{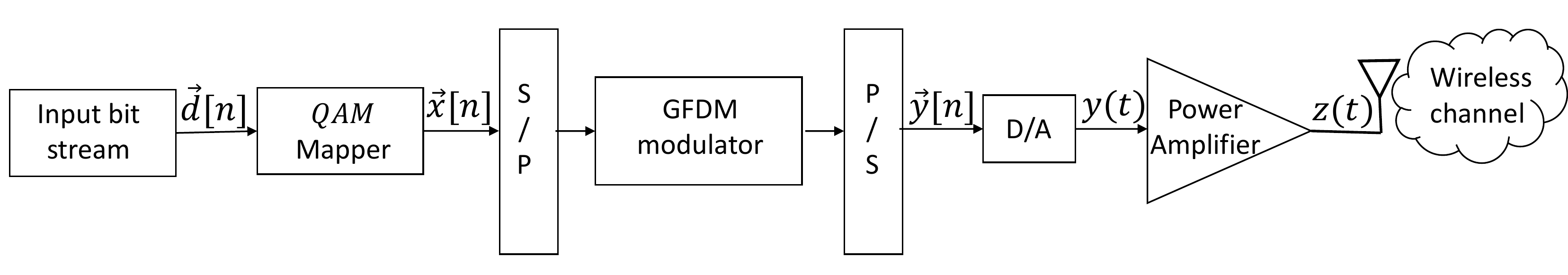}\\[-10pt]\vspace{0.5cm}
			\caption {Block diagram of transmitter }
			\label{fig:1}
		\end{center}
	\end{figure*}

	
	\section{System Model}\label{sec:model}
	
	The block diagram of typical transmitter is illustrated in Fig.~\ref{fig:1}, which contains GFDM modulator. In the first step, to generate symbols, input bit stream, $\vec{d}$, is fed into mapper, e.g. QAM with modulation order of $\mu$. Then, vector $\vec{x}$,  is converted from serial to parallel form and is passed through the GFDM modulator. Finally, the resulting vector $\vec{y}$ , converted from parallel to serial, is amplified by PA. In the next two subsections, we explain GFDM modulator and PA structures.
		\begin{figure*}[t!]
			\centering
			\includegraphics[width=1\textwidth]{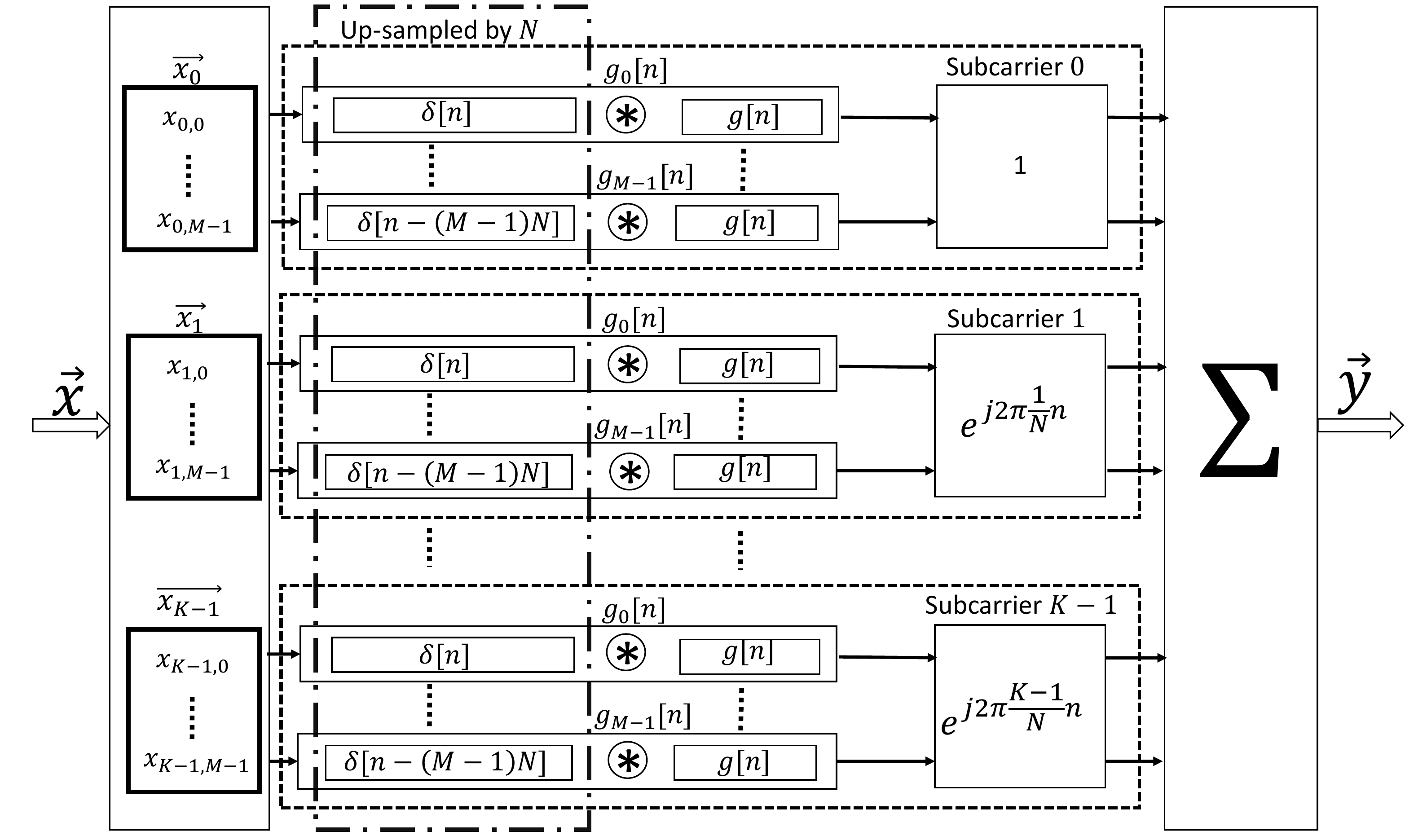}\\[-10pt]\vspace{0.5cm}
			\centering
			\caption {GFDM modulator structure}
			\label{fig:2}
		\end{figure*}
	\subsection{GFDM modulator structure}
	
	 The GFDM modulator structure is shown in Fig.~\ref{fig:2} where $M$ subsymbols per block are transmitted on $K$ subcarriers. The input of GFDM modulator is vector $\vec{x}$ , contains of $M$$K$ complex data symbols, which are independent and identically distributed ($i.i.d$). Symbols of block can be formed as $\vec{x}=[[x_{0}]^{T},[x_{1}]^{T},[x_{2}]^{T},...,[x_{K-1}]^{T}]^{T}$, including $K$ vectors each has $M$ elements, $[x_{k}]=[x_{k,0},x_{k,1},x_{k,2},...,x_{k,M-1}]^{T}$ . ${x}_{k,m}$ is indicator of $m\textrm{th}$ transmitted subsymbol on the $k\textrm{th}$ subcarrier, $k=0,1,.....,K-1$ and $m=0,1,.....,M-1$.
	According to Fig.~\ref{fig:2}, data symbols are up-sampled by the factor of $N$ where $N$ is the number of samples per symbol. Resulting vector is in the form of
	
	\begin{equation}\label{equ2}
	{s}_{k}[n]=\sqrt{\alpha }\sum\limits_{m=0}^{M-1}{{{x}_{k,m}}\delta [n-mN]}  
	\end{equation}
	where $\alpha$ is power scaling factor. Vector ${s}_{k}=[{s}_{k}[0],{s}_{k}[1],{s}_{k}[2],......,{s}_{k}[MN-1]]$ is circularly convolved with vector $\vec{g}$, which holds all coefficients of the  prototype filter  with the length $M$$N$. Finally, after up-converting to the frequency of intended subcarrier, the resulting signals are summed. Due to the circular convolution characteristics, the output of GFDM modulator per frame can be written as\cite{magh24}
	
	\begin{equation}\label{equ3}
	\begin{aligned}
	{y}_{\upsilon }[n]=&\sqrt{\alpha }\sum\limits_{k=0}^{K-1}{({s}_{k}[n]\circledast g[n] ){e^{j2\pi \dfrac{(k-\dfrac{K-1}{2})}{N}n}}}\\
	&
	=\sqrt{\alpha}\sum\limits_{k=0}^{K-1}\sum\limits_{m=0}^{M-1}{x_{k,m}g{[n-mN]}_{MN}{e^{j2\pi \dfrac{(k-\dfrac{K-1}{2})}{N}n}}}\qquad\qquad 0\le n\le MN-1 \\ 
	\end{aligned} 
	\end{equation}
	where $\upsilon$ is the frame index and $\circledast$ denotes the circular convolution. By concatenating blocks, the GFDM signal can be expressed as
	
	\begin{equation}\label{equ4}
	\begin{aligned}
	& y[n]=\sqrt{\alpha }\sum\limits_{\upsilon =-\infty }^{\infty }\sum\limits_{k=0}^{K-1}\sum\limits_{m=0}^{M-1}{x}_{k,m,\upsilon}{g}_{m}[n-\upsilon MN] \ {{e}^{j2\pi \dfrac{(k-\dfrac{K-1}{2})}{N}n}} 
	\qquad -\infty \le n\le \infty  
	\end{aligned} 
	\end{equation}
	where ${g}_{m}[n]=g{{[n-mN]}_{MN}}$ is circularly shifted version of $g[n]$.
	
	
	The continuous-time version of the GFDM modulation signal can be written as
	
	\begin{equation}\label{equ7}
	\begin{aligned}
	y(t)&=\sqrt{\alpha }\sum\limits_{\upsilon =-\infty }^{\infty }\sum\limits_{k=0}^{K-1}\sum\limits_{m=0}^{M-1}{{{x}_{k,m,\upsilon }}}{{g}_{m}}(t-\upsilon {{T}_{B}}) \ {{e}^{j2\pi \dfrac{(k-\dfrac{K-1}{2})}{{{T}_{s}}}t}}
	\qquad -\infty \le t\le \infty 
	\end{aligned}
	\end{equation}
	where ${{T}_{s}}$ is the symbol duration, ${{g}_{m}}(t)$ is continuous form of ${{g}_{m}}[n]$ with the length of $M{{T}_{s}}$ and ${{T}_{B}}$ is block duration which is equal to $M{{T}_{s}}$, respectively. By considering $\frac{N}{{{T}_{s}}}$ as sampling frequency, $MN$ discrete samples in (\ref{equ3}) are obtained. Since $\frac{1}{{{T}_{s}}}$ is bandwidth allocated to each subcarrier, the total bandwidth is $\frac{K}{{{T}_{s}}}$ by considering $K$ subcarriers.
	
	\subsection{Power amplifier}
	In this paper, to investigate the performance of nonlinear PA, behavioral modeling techniques are considered. A popular technique to model the nonlinearity of narrowband and memory less PA is polynomial function as follows as\cite{magh29}
	\begin{equation}\label{equ8}
	z(t)=\sum\limits_{i=0}^{2{{N}_{p}}+1}{{{a}_{2i+1}}{{\left| y(t) \right|}^{2i}}y(t)}
	\end{equation}
where $z(t)$ is corresponding output of PA, ${a}_{2i+1}$ are complex coefficients, $y(t)$ is the baseband input signal and $2{N}_{p}+1$ is the order of nonlinearity. In-band and adjacent-band intermodulation distortions are caused only by odd terms of nonlinearity\cite{magh31}, which can be determined by single-tone complex compression characteristics\cite{magh41}. 

	\section{Power spectral density}\label{sec:bachground}
	
\begin{figure}[t]
			\centering
			\includegraphics[width=.65\textwidth]{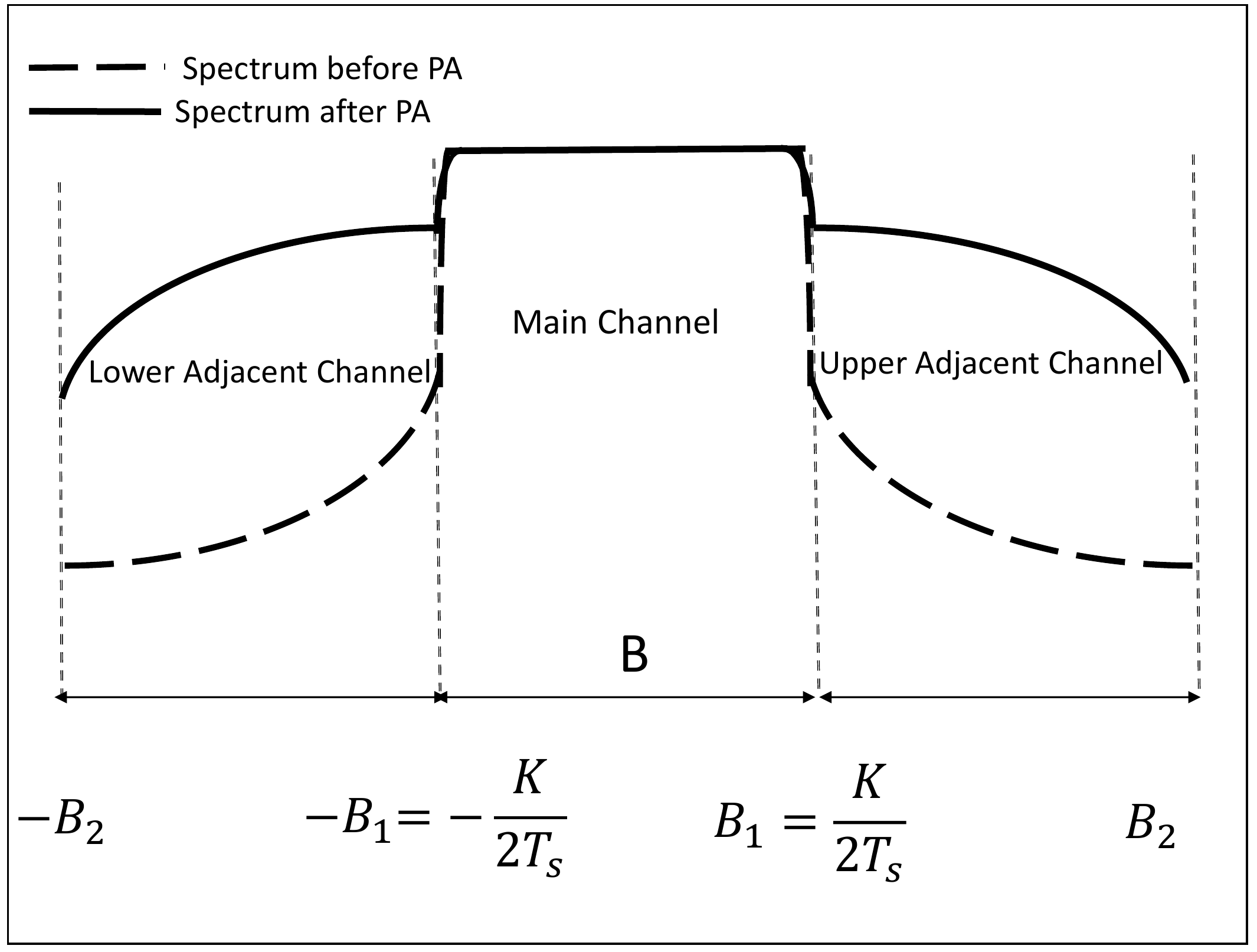}\\[-10pt]
			\centering
			\caption {The normalized PSD of the input and output nonlinear PA}
			\label{fig:3}
		\end{figure}

Nonlinear nature of PA causes out-of-band distortion. By using the polynomial modeling of PA in time domain, convolution terms appear in frequency response which result in spectral regrowth. As it is shown in Fig.~\ref{fig:3}, by considering nonlinear PA model, the spectrum of its output signal is expanded.  To investigate the effect of PA nonlinearity on OOB leakage, the spectrum of PA output signal should be calculated. Since the spectrum of output signal is undoubtedly a function of input signal, PSD of GFDM signal, as PA input, should be derived. Accordingly , in this section, PSD of GFDM is extracted and then, PSD of PA output is obtained	
		
	\subsection{PSD of GFDM modulated signal}
	
	In order to calculate the PSD of GFDM modulated signal, autocorrelation function of transmitted baseband GFDM signal, $y(t)$, should be derived. By using (\ref{equ7}), the autocorrelation function is written as follows
	
	\begin{equation}\label{equ9}
\begin{aligned}
 {{R}_{yy}}(t,\tau )&=E[y(t){{y}^{*}}(t-\tau )] \\ 
 =&\alpha \sum\limits_{{{\upsilon }_{1}}=-\infty }^{\infty }{\sum\limits_{{{\upsilon }_{2}}=-\infty }^{\infty }{\sum\limits_{{{k}_{1}}=0}^{K-1}{\sum\limits_{{{k}_{2}}=0}^{K-1}{\sum\limits_{{{m}_{1}}=0}^{M-1}{\sum\limits_{{{m}_{2}}=0}^{M-1}{E[{{x}_{{{k}_{1}},{{m}_{1}},{{\upsilon }_{1}}}}x_{{{k}_{2}},{{m}_{2}},{{\upsilon }_{2}}}^{*}]}}}}}}{{g}_{{{m}_{1}}}}(t-{{\upsilon }_{1}}{{T}_{B}})g_{{{m}_{2}}}^{*}(t-\tau -{{\upsilon }_{2}}{{T}_{B}}) \\ 
& \times {{e}^{j2\pi \dfrac{({{k}_{1}}-\dfrac{K-1}{2})}{{{T}_{s}}}t}}{{e}^{-j2\pi \dfrac{({{k}_{2}}-\dfrac{K-1}{2})}{{{T}_{s}}}(t-\tau )}} \\ 
\end{aligned}
	\end{equation}
where data symbol ${{x}_{k,m,\upsilon }}$ is independent and identically distributed random variable. Thus, autocorrelation function of ${{x}_{k,m,\upsilon }}$ can be shown as
	
	\begin{equation}\label{equ10}
E[{{x}_{{{k}_{1}},{{m}_{1}},{{\upsilon }_{1}}}}x_{{{k}_{2}},{{m}_{2}},{{\upsilon }_{2}}}^{*}]={{\overline{p}}_{x}}\delta ({{k}_{1}}-{{k}_{2}})\delta ({{m}_{1}}-{{m}_{2}})\delta ({{\upsilon }_{1}}-{{\upsilon }_{2}})
	\end{equation}
	where ${{\overline{p}}_{x}}$ is the average power of data symbols. In this paper, QAM modulation of order $\mu$ is considered which its average power is equal to $\overline{{{P}_{x}}}=\frac{2((2^{\mu} )-1)}{3}$ \cite{magh43}. By considering (\ref{equ10}), the equation (\ref{equ9}) has nonzero value for  ${{k}_{1}}={{k}_{2}}=k$ , ${{m}_{1}}={{m}_{2}}=m$ and ${{\upsilon }_{1}}={{\upsilon }_{2}}=\upsilon $, thus
	
	\begin{equation}\label{equ11}
{{R}_{yy}}(t,\tau )=\alpha {{\overline{p}}_{x}}\sum\limits_{\upsilon =-\infty }^{\infty }{\sum\limits_{k=0}^{K-1}{\sum\limits_{m=0}^{M-1}{{{g}_{m}}(t-\upsilon {{T}_{B}})g_{m}^{*}(t-\tau -\upsilon {{T}_{B}}){{e}^{j2\pi \dfrac{(k-\dfrac{K-1}{2})}{{{T}_{s}}}\tau }}}}}.
	\end{equation}
	
	According to (\ref{equ11}), autocorrelation function is time dependent and the signal is not stationary. However, ${{R}_{yy}}(t,\tau )$is periodic by ${{T}_{B}}=M{{T}_{s}}$ , which means that $y(t)$ is  cyclostationary with period $M{{T}_{s}}$. Time dependency of ${{R}_{yy}}(t,\tau )$ can be omitted by calculating average of autocorrelation function over a period  as follows
	
	\begin{equation}\label{equ12}
\begin{aligned}
 {{\overline{R}}_{yy}}(\tau )&=\frac{\alpha {{\overline{p}}_{x}}}{M{{T}_{s}}}\int\limits_{0}^{M{{T}_{s}}}{\sum\limits_{\upsilon =-\infty }^{\infty }{\sum\limits_{k=0}^{K-1}{\sum\limits_{m=0}^{M-1}{{{g}_{m}}(t-\upsilon {{T}_{B}})g_{m}^{*}(t-\tau -\upsilon {{T}_{B}}){{e}^{j2\pi \dfrac{(k-\dfrac{K-1}{2})}{{{T}_{s}}}\tau }}}}}}dt \\ 
& =\frac{\alpha {{\overline{p}}_{x}}}{M{{T}_{s}}}{{\overline{R}}_{GG}}(\tau )\sum\limits_{k=0}^{K-1}{{{e}^{j2\pi \dfrac{(k-\dfrac{K-1}{2})}{{{T}_{s}}}\tau }}}. \\ 
\end{aligned}
	\end{equation}

	By using linear convolution formula and considering ${g}_{m}(t)$  as a real term , $\overline{R}_{GG}(\tau )$can be found as
	
	\begin{equation}\label{equ13}
\begin{aligned}
 {{\overline{R}}_{GG}}(\tau )&=\int\limits_{0}^{M{{T}_{s}}}{\sum\limits_{\upsilon =-\infty }^{\infty }{\sum\limits_{m=0}^{M-1}{{{g}_{m}}(t-\upsilon {{T}_{B}})g_{m}^{*}(t-\tau -\upsilon {{T}_{B}})dt}}} \\ 
& =\sum\limits_{m=0}^{M-1}{\int\limits_{-\infty }^{\infty }{{{g}_{m}}(t)g_{m}^{*}(t-\tau )}}dt=\sum\limits_{m=0}^{M-1}{{{g}_{m}}(\tau )}\otimes {{g}_{m}}(-\tau ) \\ 
\end{aligned}
	\end{equation}
	where $\otimes$ denotes the linear convolution. By taking FT (Fourier Transform) of $\bar{R}_{yy}(\tau )$ to obtain the PSD of $y(t)$, we have
	
	\begin{equation}\label{equ14}
\begin{aligned}
 {{S}_{yy}}(f)&=\int\limits_{-\infty }^{\infty }{{{\overline{R}}_{yy}}(\tau ){{e}^{-j2\pi f\tau }}}d\tau =\frac{\alpha {{\overline{p}}_{x}}}{M{{T}_{s}}}\sum\limits_{k=0}^{K-1}{\int\limits_{-\infty }^{\infty }{{{\overline{R}}_{GG}}(\tau )}}{{e}^{-j2\pi (f-\dfrac{(k-\dfrac{K-1}{2})}{{{T}_{s}}})\tau }}d\tau  \\ 
& =\frac{\alpha {{\overline{p}}_{x}}}{M{{T}_{s}}}\sum\limits_{k=0}^{K-1}{{{S}_{GG}}(f-}\dfrac{(k-\dfrac{K-1}{2})}{{{T}_{s}}}) \\ 
\end{aligned}
	\end{equation}
	where ${{S}_{GG}}(f)$ is
	\begin{equation}\label{equ15}
\begin{aligned}
 {{S}_{GG}}(f)&=\int\limits_{-\infty }^{\infty }{{{\overline{R}}_{GG}}(\tau ){{e}^{-j2\pi f\tau }}}d\tau =\int\limits_{-\infty }^{\infty }{\sum\limits_{m=0}^{M-1}{({{g}_{m}}(\tau )}\otimes {{g}_{m}}(-\tau )){{e}^{-j2\pi f\tau }}}d\tau  \\ 
& =\sum\limits_{m=0}^{M-1}{\int\limits_{-\infty }^{\infty }{({{g}_{m}}(\tau )\otimes {{g}_{m}}(-\tau )){{e}^{-j2\pi f\tau }}}}d\tau =\sum\limits_{m=0}^{M-1}{{{G}_{m}}(f)G_{m}^{*}(-f)=}\sum\limits_{m=0}^{M-1}{{{\left| {{G}_{m}}(f) \right|}^{2}}}. \\ 
\end{aligned}
	\end{equation}
	
	By confining signal in time, frequency expansion occurs. Even though, a pulse shaped filter such as raised cosine is used, the limited power leaks into adjacent bands. In order to estimate amount of energy emission in out-of-band frequency interval, $OOB$ as a ratio between the amount of energy in out-of-band ($OB$) and in-band ($IB$) frequency range is defined as\cite{magh15}

	\begin{equation}\label{equ16}
	OOB=\frac{\left| IB \right|}{\left| OB \right|}.\frac{\int\limits_{f\in OB}{{{S}_{yy}}(f)}df}{\int\limits_{f\in IB}{{{S}_{yy}}(f)}df}.
	\end{equation}
	
	\subsection{PSD of PA output signal}
	
	PSD of PA output can be calculated by finding its autocorrelation function. According to  (\ref{equ8}), the autocorrelation function of PA output is as follows
	\begin{equation}\label{equ18}
\begin{aligned}
 {{R}_{zz}}(t,\tau )=&E[z(t){{z}^{*}}(t-\tau )]=E[\sum\limits_{{{i}_{1}}=0}^{2{{N}_{p}}+1}{{{a}_{2{{i}_{1}}+1}}{{\left| y(t) \right|}^{2{{i}_{1}}}}y(t)}\times \sum\limits_{{{i}_{2}}=0}^{2{{N}_{p}}+1}{a_{2{{i}_{2}}+1}^{*}{{\left| y(t-\tau ) \right|}^{2{{i}_{2}}}}{{y}^{*}}(t-\tau )}] \\ 
& =\sum\limits_{{{i}_{1}}=0}^{2{{N}_{p}}+1}{\sum\limits_{{{i}_{2}}=0}^{2{{N}_{p}}+1}{{{a}_{2{{i}_{1}}+1}}a_{2{{i}_{2}}+1}^{*}{{\phi }_{{{i}_{1}},{{i}_{2}}}}(t,\tau )}} \\ 
\end{aligned}
	\end{equation}
	where ${{\phi }_{{{i}_{1}},{{i}_{2}}}}(t,\tau )=E[y{{(t)}^{{{i}_{1}}+1}}y{{(t-\tau )}^{{{i}_{2}}}}{{({{y}^{*}}(t))}^{{{i}_{1}}}}{{({{y}^{*}}(t-\tau ))}^{{{i}_{2}}+1}}]$.

	By considering (\ref{equ4}) and (\ref{equ7}), $y(t)$ is continuous time version of $y[n]$ which is derived from summation of $M$$N$ independent, identically distributed random variables. Due to central limit theorem \cite{magh43} as $M$$N$ gets large, the distributions of $y(t)$ becomes Gaussian. By utilizing moments of complex Gaussian random variable $y(t)$, ${{\phi }_{{{i}_{1}},{{i}_{2}}}}(t,\tau )$ can be derived as 
\begin{equation}\label{equ21}
\begin{aligned}
 {{\phi }_{{{i}_{1}},{{i}_{2}}}}(t,\tau )=&\sum\limits_{p=0}^{\min ({{i}_{1}},{{i}_{2}})}{\left( \begin{aligned}
	& {{i}_{2}}+1 \\ 
	& p+1 \\ 
	\end{aligned} \right)\left( \begin{aligned}
	& {{i}_{1}}+1 \\ 
	& p+1 \\ 
	\end{aligned} \right)\left( \begin{aligned}
	& {{i}_{2}} \\ 
	& p \\ 
	\end{aligned} \right)\left( \begin{aligned}
	& {{i}_{1}} \\ 
	& p \\ 
	\end{aligned} \right)(p+1)!(p)!({{i}_{2}}-p)!({{i}_{1}}-p)!} \\ 
& {{({{R}_{yy}}(t,\tau ))}^{p+1}}{{(R_{yy}^{*}(t,\tau ))}^{p}}{{({{R}_{yy}}(t,0))}^{{{i}_{1}}+{{i}_{2}}-2p}} \\ 
\end{aligned}
	\end{equation}
 where, $\left( \begin{aligned}& i \\ & p \\ 
 \end{aligned} \right)=\frac{i!}{p!i-p!}$ 
 . The detail of (\ref{equ21}) is expressed in Appendix A. Based on (\ref{equ11}), (\ref{equ18}) and (\ref{equ21}), the autocorrelation function of PA output, ${{R}_{zz}}(t,\tau )$, can be derived.

 Due to the fact that ${{R}_{yy}}(t,\tau )={{R}_{yy}}(t+M{{T}_{s}},\tau )$, ${\phi_{i_{1},i_{2}}}(t,\tau )$	is periodic by $MT_{s}$ in (\ref{equ21}). Thus, in (\ref{equ18}), it is clear that ${{R}_{zz}}(t,\tau )={{R}_{zz}}(t+M{{T}_{s}},\tau )$, which shows that $z(t)$ is cyclostationary. Thus, by using (\ref{equ18}),  the average of PA output autocorrelation function can be calculated as 
 	
	\begin{equation}\label{equ51}
\begin{aligned}
	 \overline{R}_{zz}(\tau )=&\frac{1}{M{{T}_{s}}}\int\limits_{0}^{M{{T}_{s}}}{{{R}_{zz}}(t,\tau )}dt=\frac{1}{M{{T}_{s}}}\int\limits_{0}^{M{{T}_{s}}}{\sum\limits_{{{i}_{1}}=0}^{2{{N}_{p}}+1}{\sum\limits_{{{i}_{2}}=0}^{2{{N}_{p}}+1}{{{a}_{2{{i}_{1}}+1}}a_{2{{i}_{2}}+1}^{*}{{\phi }_{{{i}_{1}},{{i}_{2}}}}(t,\tau )}}}dt \\ 
	& =\sum\limits_{{{i}_{1}}=0}^{2{{N}_{p}}+1}{\sum\limits_{{{i}_{2}}=0}^{2{{N}_{p}}+1}{{{a}_{2{{i}_{1}}+1}}a_{2{{i}_{2}}+1}^{*}{{\phi }_{{{i}_{1}},{{i}_{2}}}}(\tau )}} \\ 
\end{aligned}
\end{equation}
	where ${{\phi }_{{{i}_{1}},{{i}_{2}}}}(\tau )=\frac{1}{M{{T}_{s}}}\int\limits_{0}^{M{{T}_{s}}}{{{\phi }_{{{i}_{1}},{{i}_{2}}}}(t,\tau )}dt.$ Finally, by taking FT (Fourier Transform) of (\ref{equ51}), the PSD of $z(t)$ can be calculated as
	
	\begin{equation}\label{equ92}
\begin{aligned}
 {{S}_{zz}}(f)=&\int\limits_{-\infty }^{\infty }{\overline{R}_{zz}(\tau ){{e}^{-j2\pi f\tau }}}d\tau =\int\limits_{-\infty }^{\infty }{\sum\limits_{{{i}_{1}}=0}^{2{{N}_{p}}+1}{\sum\limits_{{{i}_{2}}=0}^{2{{N}_{p}}+1}{{{a}_{2{{i}_{1}}+1}}a_{2{{i}_{2}}+1}^{*}{{\phi }_{{{i}_{1}},{{i}_{2}}}}(\tau ){{e}^{-j2\pi f\tau }}}}}d\tau  \\ 
& =\sum\limits_{{{i}_{1}}=0}^{2{{N}_{p}}+1}{\sum\limits_{{{i}_{2}}=0}^{2{{N}_{p}}+1}{{{a}_{2{{i}_{1}}+1}}a_{2{{i}_{2}}+1}^{*}{{\phi }_{{{i}_{1}},{{i}_{2}}}}(f)}}. \\ 
\end{aligned}
	\end{equation}

	By considering ${{\phi }_{{{i}_{1}},{{i}_{2}}}}(f)=\int\limits_{-\infty }^{\infty }{{{\phi }_{{{i}_{1}},{{i}_{2}}}}(\tau ){{e}^{-j2\pi f\tau }}}d\tau$, the PSD of PA output is derived as

 \begin{equation}\label{equ57}
 \begin{aligned}
 & {{S}_{zz}}(f)=\sum\limits_{{{i}_{1}}=0}^{2{{N}_{p}}+1}{\sum\limits_{{{i}_{2}}=0}^{2{{N}_{p}}+1}{{{a}_{2{{i}_{1}}+1}}a_{2{{i}_{2}}+1}^{*}}} \\ 
 & \sum\limits_{p=0}^{\min ({{i}_{1}},{{i}_{2}})}{\left( \begin{aligned}
 	& {{i}_{2}}+1 \\ 
 	& p+1 \\ 
 	\end{aligned} \right)\left( \begin{aligned}
 	& {{i}_{1}}+1 \\ 
 	& p+1 \\ 
 	\end{aligned} \right)\left( \begin{aligned}
 	& {{i}_{2}} \\ 
 	& p \\ 
 	\end{aligned} \right)\left( \begin{aligned}
 	& {{i}_{1}} \\ 
 	& p \\ 
 	\end{aligned} \right)(p+1)!(p)!({{i}_{2}}-p)!({{i}_{1}}-p)!}{{(\alpha {{\overline{p}}_{x}})}^{{{i}_{1}}+{{i}_{2}}+1}}{{(K)}^{{{i}_{1}}+{{i}_{2}}-2p}}\frac{1}{M{{T}_{s}}} \\ 
 & {{P}_{{{i}_{1}},{{i}_{2}}}}(f)\otimes \left( \sum\limits_{{{k}_{1}},...,{{k}_{p+1}}=0}^{K-1}\quad{\sum\limits_{k_{1}^{'},...,k_{p}^{'}=0}^{K-1}{\delta (f-\frac{(\sum\limits_{j=1}^{p+1}{{{k}_{j}}}-\sum\limits_{{{j}^{'}}=1}^{p}{{{k}_{{{j}^{'}}}}-\frac{K-1}{2}})}{{{T}_{s}}})}} \right) \\ 
 \end{aligned}
	\end{equation}	 
where ${{P}_{{{i}_{1}},{{i}_{2}}}}(f)$ is equal to
	
 \begin{equation}\label{equ90}
\begin{aligned}
 {{P}_{{{i}_{1}},{{i}_{2}}}}(f)=&\sum\limits_{{{m}_{1}},...,{{m}_{p+1}}=0}^{M-1}{\sum\limits_{m_{1}^{'},...,m_{p}^{'}=0}^{M-1}{\sum\limits_{m_{1}^{''},...,m_{{{i}_{1}}+{{i}_{2}}-2p}^{''}=0}^{M-1}{\left( {{G}_{{{m}_{1}}}}(f),...\otimes {{G}_{{{m}_{p+1}}}}(f)\otimes G_{m_{1}^{'}}^{*}(-f) \right.}}} \\ 
& \left.\otimes~ G_{m_{p}^{'}}^{*}(-f)\otimes G_{m_{1}^{''}}^{*}(-f)\otimes ...\otimes G_{m_{{{i}_{1}}+{{i}_{2}}-2p}^{''}}^{*}(-f)\otimes {{G}_{m_{1}^{''}}}(f)\otimes ...\otimes {{G}_{m_{{{i}_{1}}+{{i}_{2}}-2p}^{''}}}(f) \right) \\ 
& \times \left( G_{_{{{m}_{1}}}}^{*}(f)\otimes ...\otimes G_{{{m}_{p+1}}}^{*}(f)\otimes {{G}_{m_{1}^{'}}}(-f)\otimes ...\otimes {{G}_{m_{p}^{'}}}(-f) \right) \\ 
\end{aligned}
 \end{equation}

 The detail of (\ref{equ57}) and (\ref{equ90}) is expressed in Appendix B. As it is clear from (\ref{equ90}), ${{P}_{{{i}_{1}},{{i}_{2}}}}(f)$ only depends on filters which are extracted from prototype filter. Indeed, (\ref{equ57}) is a closed form generic function for PSD of PA output signal. 	
	
	\section{ Nonlinear Parameter Extraction}\label{sec:1db}
	
 A suitable operating point of a nonlinear PA can be selected by considering $1\textrm{dB}$ compression point and saturation point. The point in which the output power of a nonlinear PA is $1\textrm{dB}$ below the linear one, is known as $1\textrm{dB}$ compression point. Also, saturation point specifies the saturation region of PA where increase in input power causes no further rise in output power. On the other hand, due to nonlinear behavior of PA, the spectrum of transmitted signal can leak into adjacent frequency bands. By calculating ACP, this disruptive effect caused by nonlinear PA can be predicted for any input power scaling factor. The ratio between ACP and the power in the main channel is called ACPR. In this section, first, a closed-form expression of $1\textrm{dB}$ compression point and saturation point are derived. Then, ACP and ACP metrics are considered.
	
	\subsection{$1\textrm{dB}$ compression point and saturation point}
    In order to calculate the $1\textrm{dB}$ compression point, the total nonlinear PA output power should be obtained as a function of input power scaling factor. According to (\ref{equ51}), the total output power can be expressed as
   
   \begin{equation}\label{equ26}
{{P}_{z}(\alpha)}={{\overline{R}}_{zz}}(0)=\sum\limits_{{{i}_{1}}=0}^{2{{N}_{p}}+1}{\sum\limits_{{{i}_{2}}=0}^{2{{N}_{p}}+1}{{{a}_{2{{i}_{1}}+1}}{{a}^{*}}_{2{{i}_{2}}+1}{{\phi }_{{{i}_{1}},{{i}_{2}}}}(0)}}
    \end{equation}
	where 
	   \begin{equation}\label{equ40}
\begin{aligned}
 {{\phi }_{{{i}_{1}},{{i}_{2}}}}(0)=&\frac{1}{M{{T}_{s}}}\sum\limits_{p=0}^{\min ({{i}_{1}},{{i}_{2}})}{\left( \begin{aligned}
	& {{i}_{2}}+1 \\ 
	& p+1 \\ 
	\end{aligned} \right)\left( \begin{aligned}
	& {{i}_{1}}+1 \\ 
	& p+1 \\ 
	\end{aligned} \right)\left( \begin{aligned}
	& {{i}_{2}} \\ 
	& p \\ 
	\end{aligned} \right)\left( \begin{aligned}
	& {{i}_{1}} \\ 
	& p \\ 
	\end{aligned} \right)(p+1)!(p)!({{i}_{2}}-p)!({{i}_{1}}-p)!} \\ 
& \int\limits_{0}^{M{{T}_{s}}}{{{({{R}_{yy}}(t,0))}^{{{i}_{1}}+{{i}_{2}}+1}}}dt.\\ 
\end{aligned}
	   \end{equation}

	Note that ${{R}_{yy}}(t,0)=\alpha K{{\overline{p}}_{x}}\sum\limits_{m=0}^{M-1}{{{\left| {{g}_{m}}(t) \right|}^{2}}}$. According to (\ref{equ11}), ${{R}_{yy}}(t,0)$ is function of $\alpha$, thus the total output power as a function of input power scaling factor can be expressed as
	
	\begin{equation}\label{equ27}
{{P}_{z}(\alpha)}=\sum\limits_{j=1}^{\infty }{{{A}_{j}}{{\alpha }^{j}}}
	\end{equation}
	where
	
	\begin{equation}\label{equ28}
 \begin{aligned}
  &{{A}_{j}}=\frac{{{\left( K{{\overline{p}}_{x}} \right)}^{j}}}{M{{T}_{s}}}\int\limits_{0}^{M{{T}_{s}}}{{\left( \sum\limits_{m=0}^{M-1}{{{\left| {{g}_{m}}(t) \right|}^{2}}} \right)}^{j}dt} \\ 
 & \times \sum\limits_{\begin{smallmatrix} 
 	{{i}_{1}}+{{i}_{2}}+1=j \\ 
 	{{i}_{1}}+{{i}_{2}}\ge 0 
 	\end{smallmatrix}}{{{a}_{2{{i}_{1}}+1}}a_{2{{i}_{2}}+1}^{*}\sum\limits_{p=0}^{\min ({{i}_{1}},{{i}_{2}})}{\left( \begin{aligned}
 		& {{i}_{2}}+1 \\ 
 		& p+1 \\ 
 		\end{aligned} \right)\left( \begin{aligned}
 		& {{i}_{1}}+1 \\ 
 		& p+1 \\ 
 		\end{aligned} \right)\left( \begin{aligned}
 		& {{i}_{2}} \\ 
 		& p \\ 
 		\end{aligned} \right)\left( \begin{aligned}
 		& {{i}_{1}} \\ 
 		& p \\ 
 		\end{aligned} \right)(p+1)!(p)!({{i}_{2}}-p)!({{i}_{1}}-p)!}}. \\ 
 \end{aligned}
	\end{equation}

	The saturation point is the smallest positive root of derivative of ${{P}_{z}(\alpha)}$ with respect to $\alpha$ which is derived as $\frac{\partial }{\partial z}{{P}_{z}}=\sum\limits_{j=1}^{\infty }{j{{A}_{j}}{{\alpha }^{j-1}}}=0$. Similarly, $1dB$ compression point is the smallest positive root of $10{{\log }_{10}}({{P}_{z}}(\alpha ))=10{{\log }_{10}}({{A}_{1}}\alpha )-1$ according to definition on it.  
	
	
	
	\subsection{ACP and ACPR}
	
	As mentioned, ACP determines the amount of power in adjacent channel. By considering (\ref{equ57}), PSD of PA output signal is obtained by summation of some terms which not only have the same bandwidth as GFDM signal ($-{{B}_{1}}\le f\le {{B}_{1}}$), but also have multiples of bandwidth of the main channel, which causes spectral regrowth. Therefore, the bandwidth of PA output signal is more than the input signal bandwidth. As a result, the ACP can be derived by integrating over the bandwidth of upper or lower adjacent channel. In general, upper ACP can be obtained by
	
	\begin{equation}\label{equ31}
	ACP(p )=\int\limits_{{{B}_{1}}}^{{{B}_{2}}}{{{S}_{zz}}(f)df}.
	\end{equation}
	
	As can be seen in Fig.~\ref{fig:3} , $\left[ {{B}_{1}},{{B}_{2}} \right]$ is a frequency interval of upper adjacent channel. 
	Moreover, ratio between power of adjacent channel to the power of main channel, ACPR, is expressed as
	
	\begin{equation}\label{equ32}
	ACPR(p )=\frac{ACP(p )}{\int\limits_{-{{B}_{1}}}^{{{B}_{1}}}{{{S}_{zz}}(f)df}}
	\end{equation}
	where $\left[ -{{B}_{1}},{{B}_{1}} \right]$ is a frequency interval of the main channel. 

				\begin{table}[t]
					\caption{\small Parameters of GFDM modulation and Nonlinear coefficients of the PA model.}
					\centering
					\begin{tabular}{ |c|c|  }
						\hline
						Parameter & Value \\
						\hline
						Mapping& $16-QAM$ \\
						\hline
						Filter type& $Raised-cosine$ \\
						\hline
						Roll-off factor& $0.3$ \\
						\hline
						Symbol duration ($T_{s}$)& 33.3 $\mu s$ \\
						\hline
						Number of subcarrier ($K$)& 64 \\
						\hline
						Samples per symbol ($N$)& 320 \\
						\hline
						Number of subsymbols ($M$)& 5 and 35 \\
						\hline
						Subcarrier spacing ($\frac{1}{T_{s}}$)&30 $kHz$\\
						\hline
						Signal Bandwidth ($2 B_{1}=\frac{K}{T_{s}}$) &1.92 $MHz$\\
						\hline
						Sampling frequency ($\frac{N}{T_{s}}$) & 9.6 $MHz$\\
						\hline
						Nonlinear coefficient ($a_{1}$) & 14.9740 +$j$0.0519\\
						\hline
						Nonlinear coefficient ($a_{3}$) & -23.0954 +$j$4.9680 \\
						\hline
						Nonlinear coefficient ($a_{5}$) & 21.3936 +$j$0.4305 \\
						\hline
					\end{tabular}
					\label{table}
				\end{table}
	\section{simulation and Numerical results}\label{sec:sim}
	
	In this section, derived analytical expressions are verified with simulation results. Parameters of GFDM modulation and complex coefficients of a fifth order polynomial function of PA model\cite{magh50} are represented in table \ref{table}. Monte Carlo simulation with 1000 GFDM symbols is used to generate GFDM modulated signal. To estimate the PSD, averaged periodogram algorithm with 50$\%$ overlap and hanning window is used\cite{magh46}. For extracting simulation results, the length of FFT is set to $65536$. Causal and FIR prototype filter, $g(t)$, is shifted length $\frac {M{T}_{s}}{2} $ of truncated infinite Raised cosine filter in interval $\left[ \frac{-M{{T}_{s}}}{2},\frac{M{{T}_{s}}}{2} \right]$ which is normalized to unit energy. According to $g(t)$, other filters, ${{g}_{m}}(t)$, are as follows
	
	\begin{equation}\label{equ33}
{{g}_{m}}(t)=\left\{ \begin{aligned}
& g(t-m{{T}_{s}}),~ m{{T}_{s}}<t\le M{{T}_{s}} \\ 
& g((M-m){{T}_{s}}+t),~ 0<t\le m{{T}_{s}} \\ 
\end{aligned} \right.
	\end{equation}

	Furthermore, ${{g}_{m}}[n]$ are filter coefficients which are discrete form of ${{g}_{m}}(t)$. To evaluate nonlinear behavior of PA, the PSD of PA output should be calculated. For this purpose, PSD of GFDM signal is simulated for given parameters. Then, by simulating PSD of PA output, its spectrum expansion is studied . Note that all simulation results are verified with analytic expression.
				\begin{figure*}[t]
					\begin{subfigure}[b]{0.5\textwidth}
						\includegraphics[width=1\textwidth]{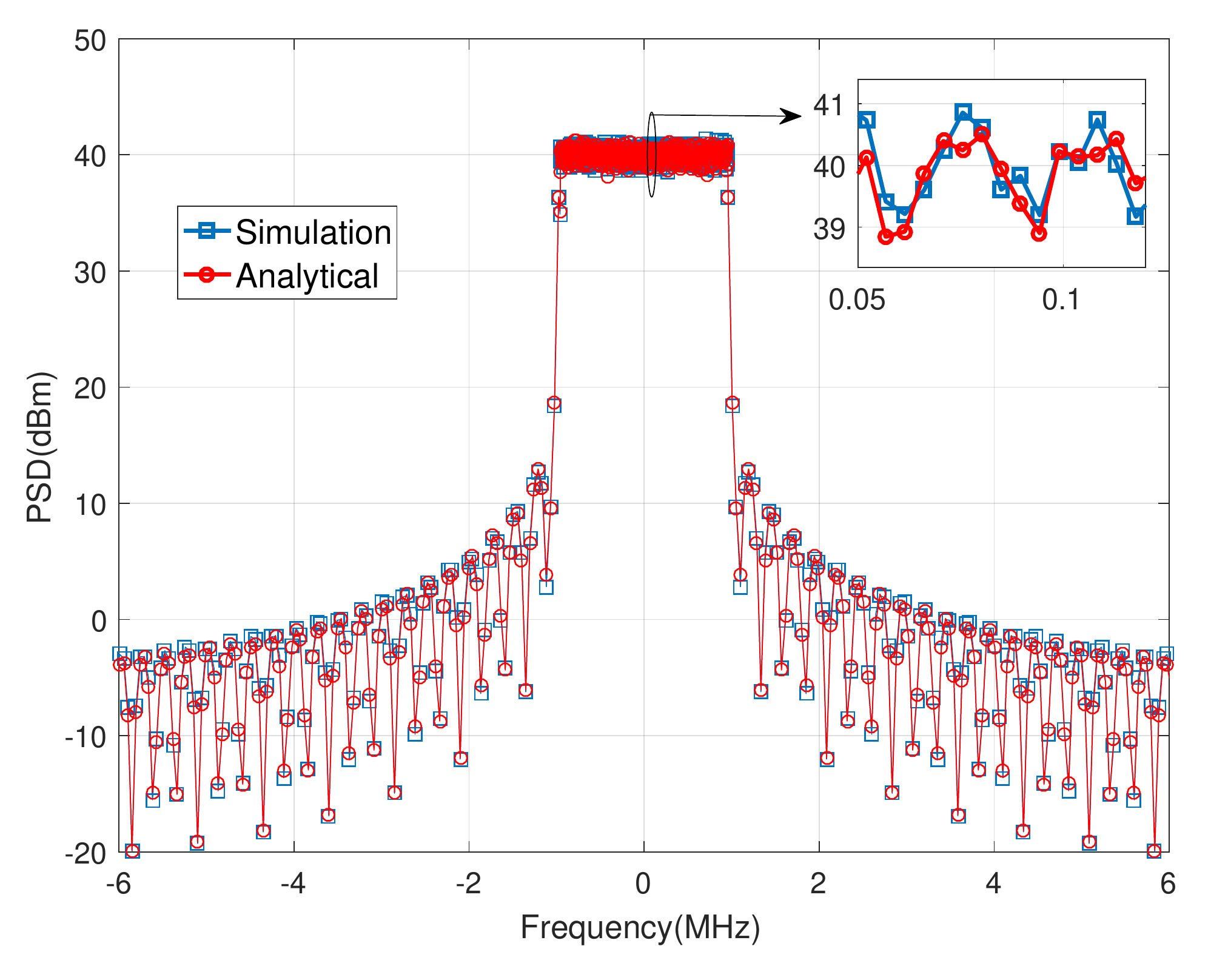}\\[-10pt]
						\centering
						\caption{\small $M=5$}
						\label{PSDGFDMM5}
					\end{subfigure} 
					\begin{subfigure}[b]{0.5\textwidth}
						\includegraphics[width=1\textwidth]{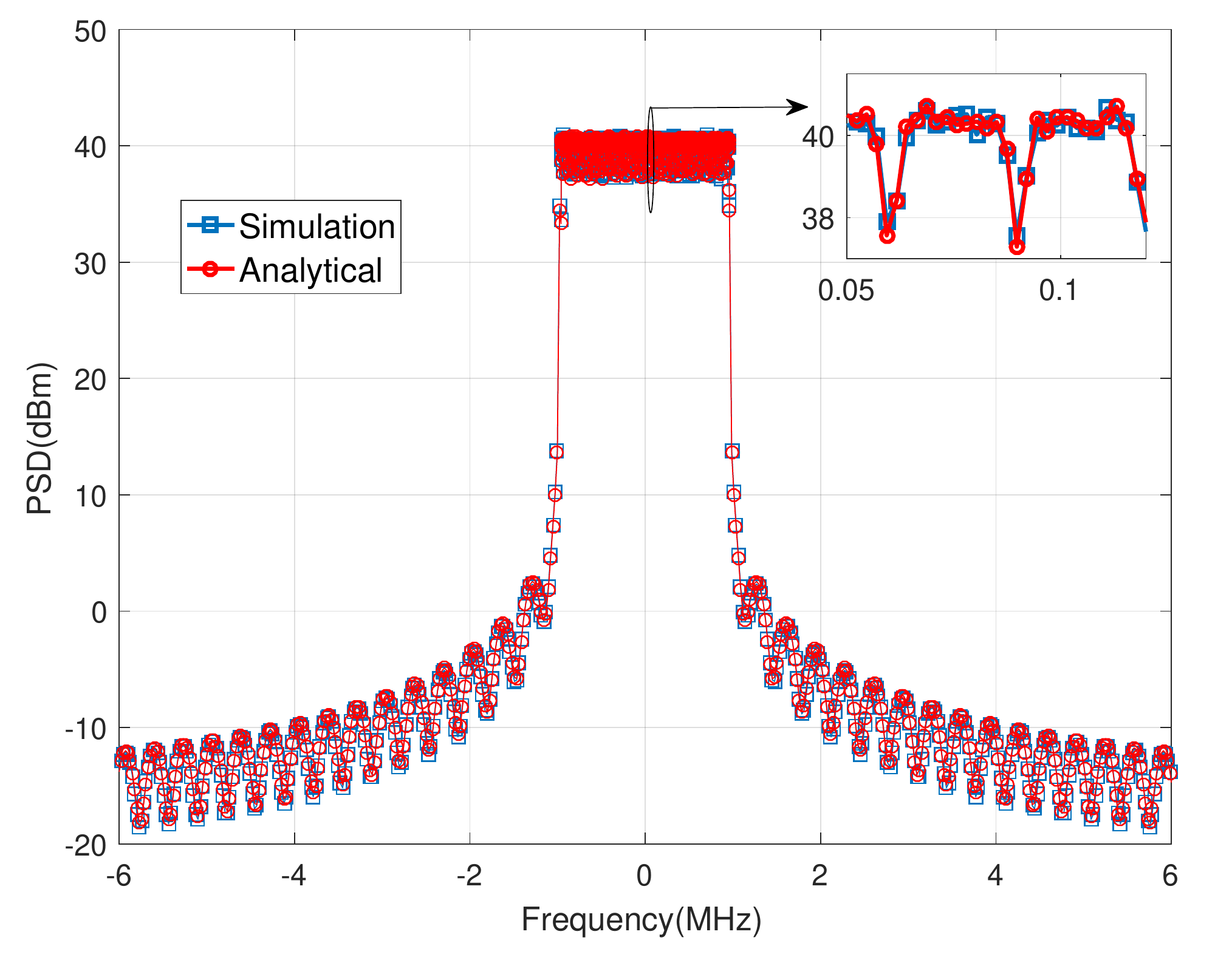}\\[-10pt]
						\centering
						\caption{\small $M=35$}
						\label{PSDGFDMM35}
					\end{subfigure}
					\caption {\small Power spectral density of GFDM modulated signal for two different of subsymbols.}
					\label{PSDGFDM}
				\end{figure*}
				
	\subsection{PSD of GFDM signal}
	In this part, PSD of GFDM signal are examined and the parameters which have effect on it, are investigated. In Fig.~\ref{PSDGFDM}, PSDs of GFDM modulated signals with $k=64$ (number of subcarriers) and two different numbers of subsymbols ($M=5$ and $M=35$) are illustrated, where the scaling factor, $p =1$ , is considered in these two figures. As can be seen, the simulation results verify the derived analytical formula for PSD of GFDM (\ref{equ14}). By comparing  Fig.~\ref{PSDGFDM}(a) and Fig.~\ref{PSDGFDM}(b), it can be concluded that difference between in-band and out-of-band values of PSD is varied for different number of subsymbols. OOB radiation for $M=35$ is approximately $6dB$ below for $M=5$. To justify this behavior, we refer to the procedure of truncating infinite filter for obtaining prototype filter which causes appearance of nonzero side lobes in frequency domain. By increasing the number of subsymbols, time span of filter  increases and the spectrum approaches to its ideal position which causes suppression of side lobs. Consequently, by increasing the number of time slots, OOB leakage of GFDM signal which is a result of side lobe reduction, is diminished.
		\begin{figure}[t]
			\centering
			\includegraphics[width=.6\textwidth]{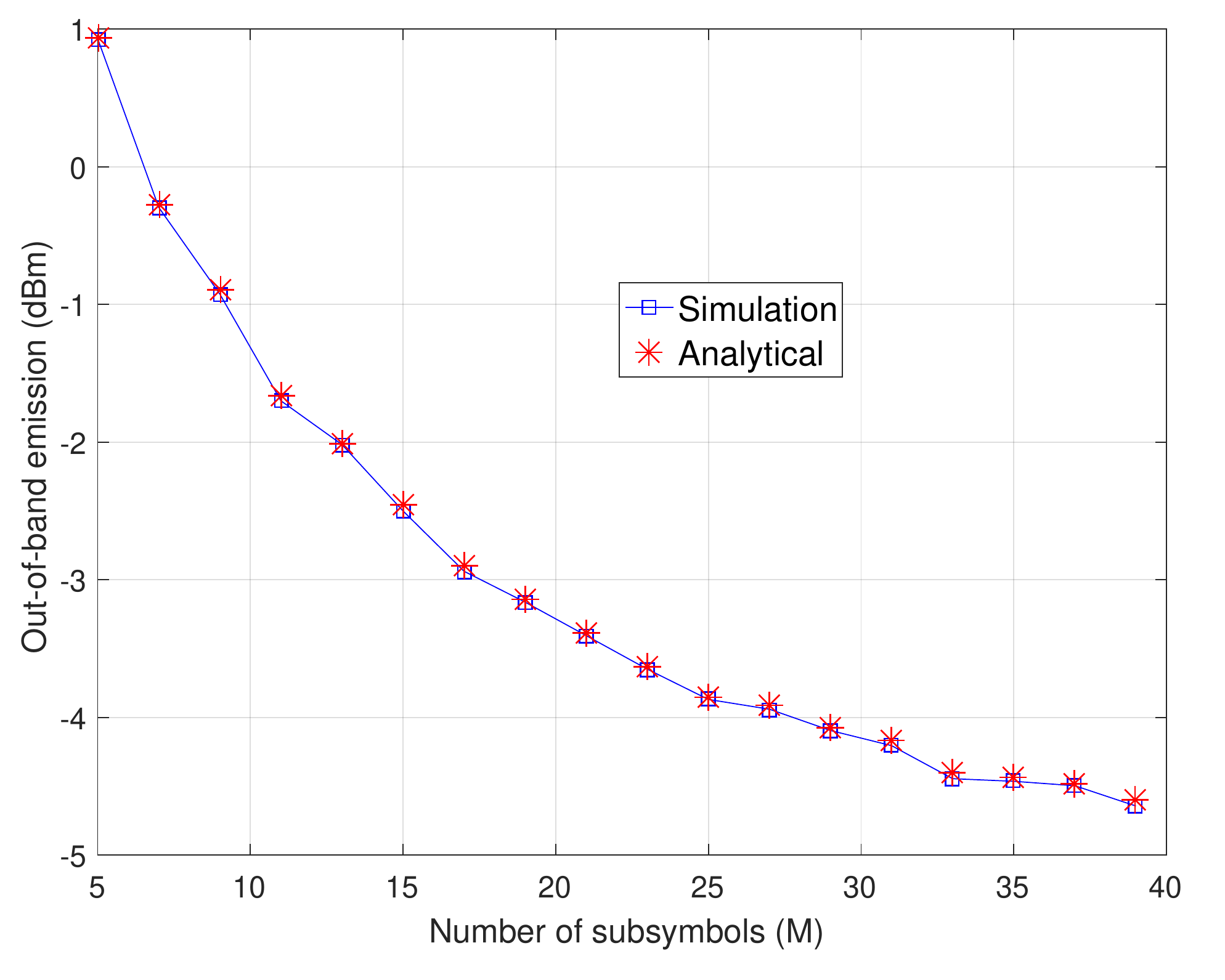}\\[-12pt]
			\caption {OOB radiation of GFDM versus different numbers of subsymbols}
			\label{OOB}
		\end{figure}
	
	In Fig.~\ref{OOB}, OOB leakages of GFDM modulated signal with 64 subcarriers versus varied number of subsymbols is represented in which the input power scaling factor is same as Fig(\ref{PSDGFDM}). In both simulation and analytic result (\ref{equ16}), frequency of $OOB$ and $B$ is  considered in the range of ${{\left[ 0.96,4.8 \right]}_{MHz}}$ and ${{\left[ -0.96,0.96 \right]}_{MHz}}$, respectively. As expected, increasing the number of subsymbols causes reduction in OOB leakage. However, the other performance metrics of system, such as bit error rate, may be degraded. Again, the simulation result verifies the derived analytical formula.

	\subsection{PSD of output PA}
	
	In pervious part, PSD of GFDM signal is investigated. In this part, the effects of PA nonlinearity on GFDM signal performance will be considered. For this purpose, first the analytical expressions of PA output power, $1\textrm{dB}$ compression and saturation points as critical points of given PA model are verified by simulation result. After that, PSD of PA output and ACP/ACPR metrics, by given analytical formulas, are studied. Similarly, two sets of GFDM, $64$ subcarriers and $5$, $35$ subsymbols, are considered to show the effect of different number of subsymbol on the results.
					\begin{figure}
						\centering
						\includegraphics[width=.6\textwidth]{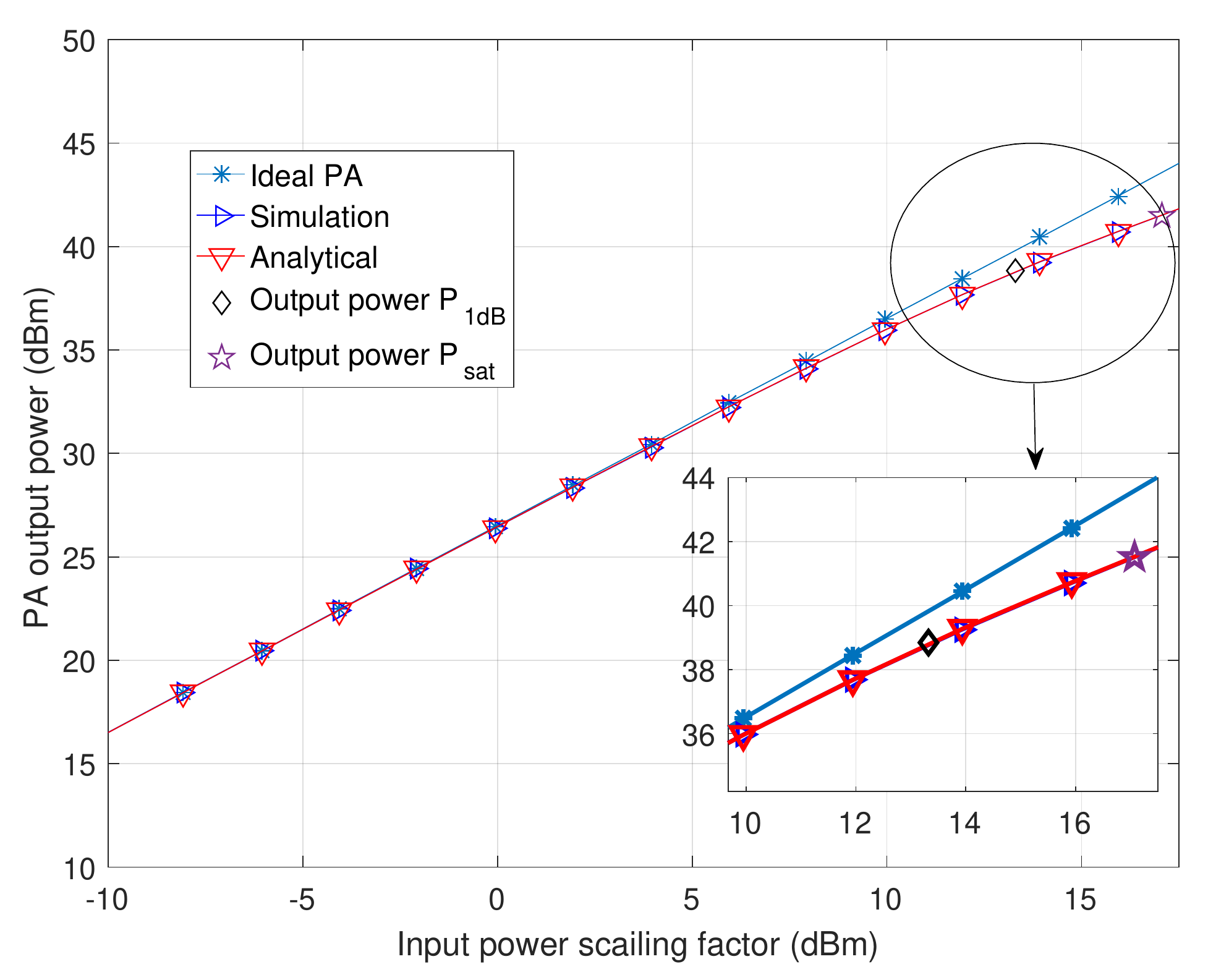}\\[-10pt]
						\caption {PA output power curves versus input power scaling factor.}
						\label{pout-pin}
					\end{figure}
			
	Fig.~\ref{pout-pin} represents $AM/AM$ characteristic of considered PA model derived by simulation and analytical formula (\ref{equ27}). Also, the $1\textrm{dB}$ compression and saturation points are ${{p}_{1dB}} =13.3 dBm$ and ${{p}_{sat}}=17.07 dBm$. 
	This figure shows that theoretical and simulation results are clearly matched with each other so the total PA output power expression in (\ref{equ27}) is verified.

							\begin{figure*}[t]
								\begin{subfigure}[b]{0.5\textwidth}
									\includegraphics[width=.98\textwidth]{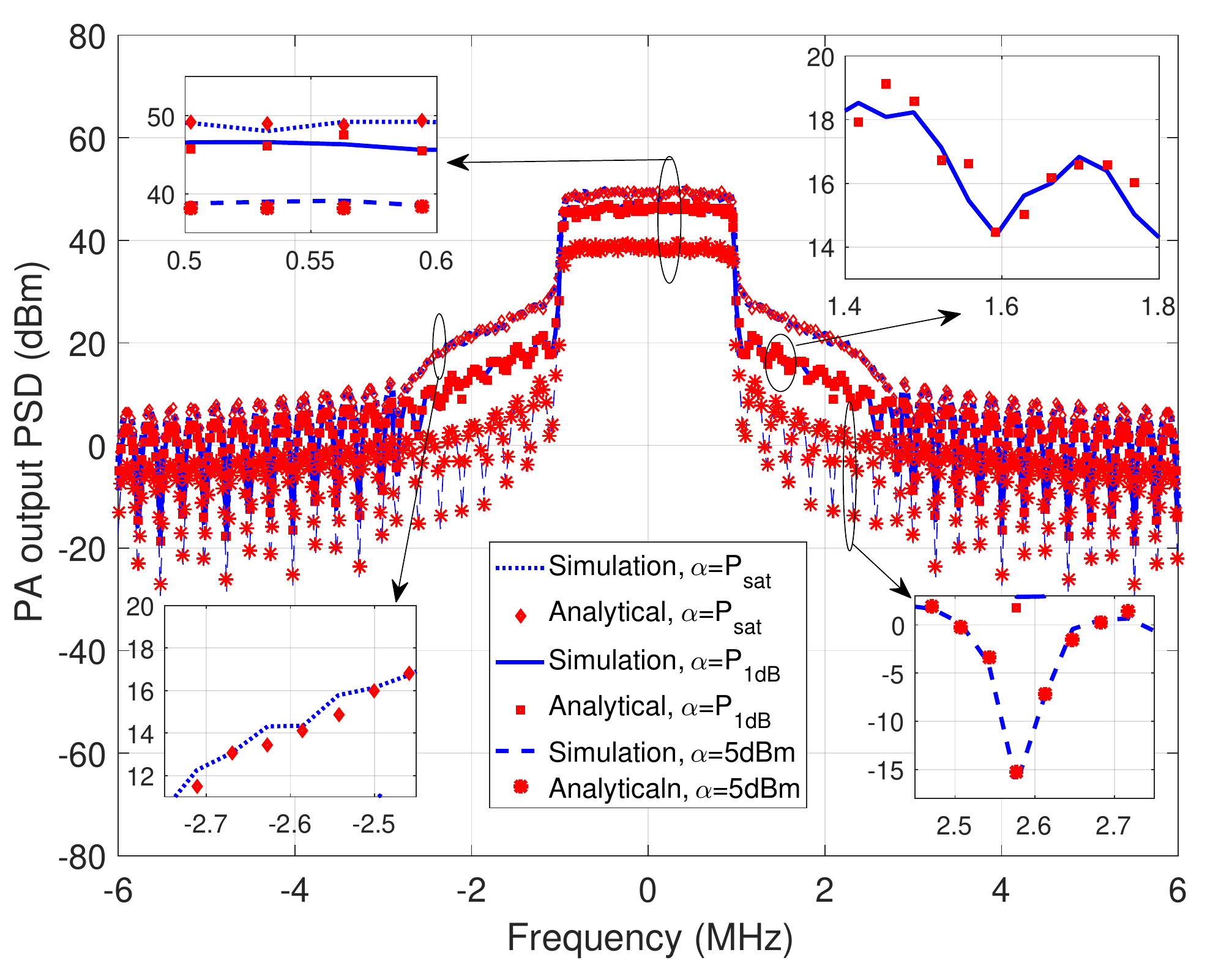}\\[-10pt]
									\centering
									\caption{\small $M=5$}
									\label{PAPSDM5}
								\end{subfigure} 
								\begin{subfigure}[b]{0.5\textwidth}
									\includegraphics[width=1\textwidth]{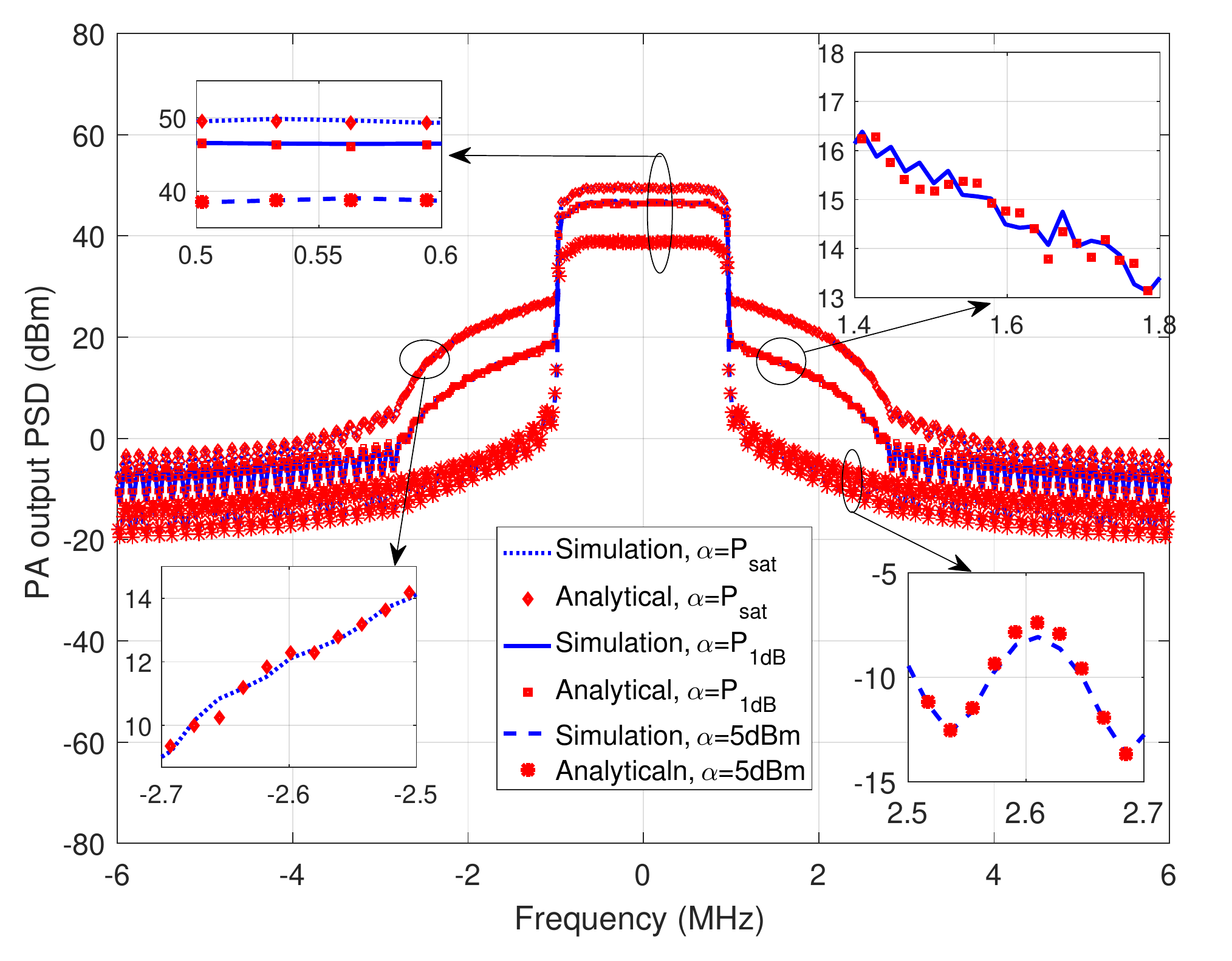}\\[-10pt]
									\centering
									\caption{\small $M=35$}
									\label{PAPSDM35}
								\end{subfigure}
								\caption {\small Power spectral density of PA output for two diffrent number of subsymbols.}
								\label{PAPSD}
							\end{figure*} 
	
		\begin{figure}[t]
			\begin{subfigure}[b]{0.5\textwidth}
				\includegraphics[width=1\textwidth]{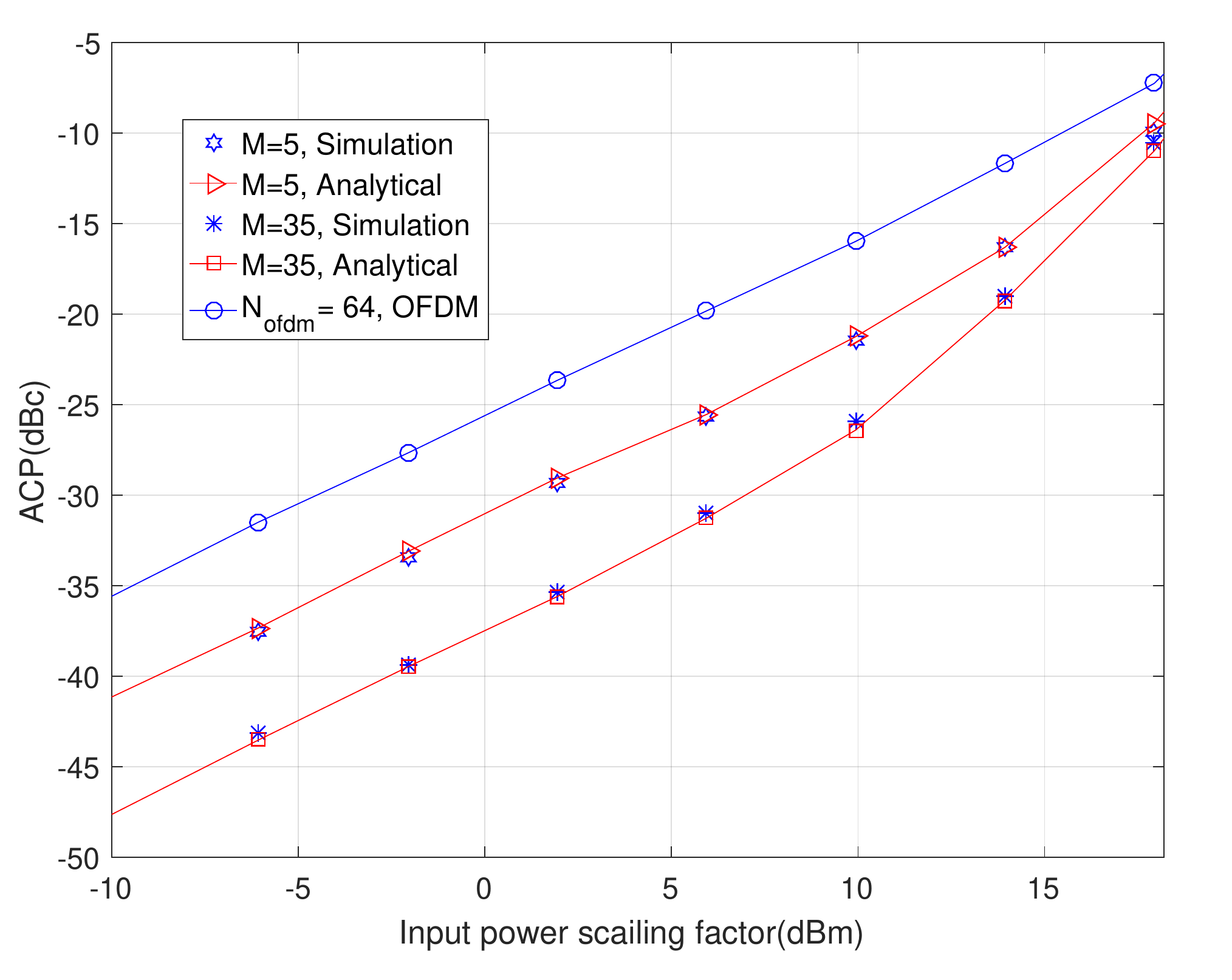}\\[-10pt]
				\centering
				\caption{}
				\label{ACP}
			\end{subfigure} 
			\begin{subfigure}[b]{0.5\textwidth}
				\includegraphics[width=1\textwidth]{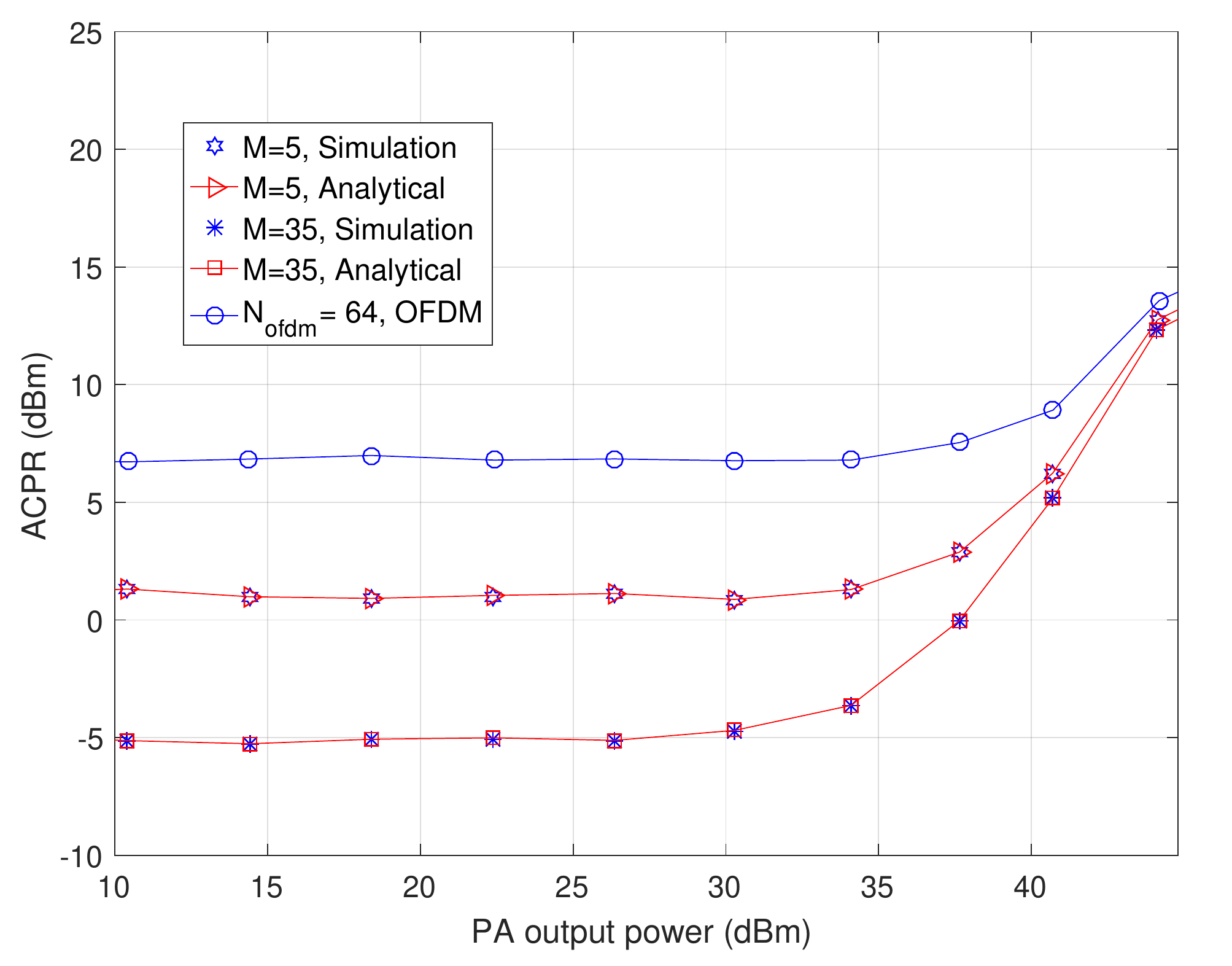}\\[-10pt]
				\centering
				\caption{}
				\label{ACPR}
			\end{subfigure}
			\caption {\small (a) ACP versus input power scaling factor and (b) ACPR versus  PA output power}
			\label{ACPACPR}
		\end{figure}

	In Fig.~\ref{PAPSD}, PSD of PA output (equation (\ref{equ57})) is compared with the simulation results. In both Fig.~\ref{PAPSD}(a) and Fig.~\ref{PAPSD}(b), PSDs of output signal of power amplifier are depicted for three input power scaling factors ${\alpha}=5 dBm$, ${\alpha}={{p}_{1dB}}$ and ${\alpha}={{p}_{sat}}$. Note that the first one is selected in linear region of PA. As expected, by increasing the input power scaling factor and approaching to the saturation point, adjacent channel power in ${{\left[ 0.96,4.8 \right]}_{MHz}}$ interval increases. As it is shown in figures, by increasing the number of subsymbols, for two power inputs, ${{p} _ {1dB}} $ and ${{p}_{sat}}$, spectral regrowth approximately stays same and the number of subsymbols does not have effect on it. But, in ${\alpha}=5 dBm$, OOB radiation in adjacent channel for $M=35$ is approximately $5dBm$ less than $M=5$.  It is due to the fact that by boosting the input power scaling factor, the impact of the nonlinear terms in the spectral regrowth, which are produced by multiple self-convolution of GFDM spectrum, becomes stronger. Since the in-band spectrum of GFDM signal is independent of $M$, these terms are independent too. For ${\alpha}=5 dBm$, PA operates in linear region and PSD of its output is just relied on the linear term, though by increasing the number of subsymbols OOB leakage reduces. Therefore, by considering ${{p}_{1dB}}$ and ${{p}_{sat}}$ as input power scaling factors, the nonlinear terms have dominant influence and same result is obtained for $M=5$ and $M=35$ due to its independency of $M$.
	
	In Fig.~\ref{ACPACPR}, simulation and analytical results of ACP and ACPR metrics are illustrated. As mentioned, ${{\left[ 0.96,4.8 \right]}_{MHz}}$ is considered as frequency interval of adjacent channel. In Fig.~\ref{ACPACPR}, theoretical expression of ACP in (\ref{equ31}) and ACPR in (\ref{equ32}) are compared with the Monte Carlo simulation results for two sets of GFDM modulations parameters to examine the accuracy of derived expressions. As expected, since PA operates in the linear region, OOB leakage is affected by number of subsymbols. By increasing the input power scaling factor and entering to the nonlinear region of PA, the spectral regrowth caused by nonlinearity is dominant and results in approaching to same value for $M=5$ and $M=35$. Moreover, ACP and ACPR of OFDM modulated signal with 64 subcarriers due to the nonlinear PA are represented in Fig.~\ref{ACPACPR}. As expected, GFDM modulated signal based on filter bank structure has better performance compared with OFDM modulated signal both in linear region and nonlinear region which shows that gain of GFDM modulated signal after PA is preserved.

	\section{Conclusion}\label{sec:conclusion}
	
	In this paper, by utilizing cyclostationary property of autocorrelation function of GFDM signal, the analytical expression of its PSD was derived. According to the derived analytical expression for PSD, OOB radiation of GFDM spectrum is  decreased by increasing the number of subsymbols which is verified by simulation results. Then, by considering the polynomial function to model the nonlinear behavior of PA, the PSD of output signal was extracted and verified by simulation results. Amplitude (AM/AM) and Phase (AM/PM) distortions  was considered with memoryless model. By representing PA output power as a function of input power and coefficients of polynomial function
	 dependence of spectral regrowth to input power and the number of subsymbols was proved. For low input power, increasing the number of subsymbols causes reduction in OOB emission though, this effect is diminished for high input power by approaching to nonlinear zone of PA. Moreover, ACP and ACPR were obtained which confirmed our aforementioned results. All analytical expressions were verified by simulation results by where a good agreement between them in all cases were obtained. Furthermore, comparison of PA output ACP and ACPR  with OFDM input and GFDM input showed that the benefit of GFDM are preserved after PA. 
	\begin{appendices}
	\section{}
Moments of complex Gaussian random variable $l$ are given by\cite{magh44}
\begin{equation}\label{equ53}
\begin{aligned}
& E[{{l}_{1}}{{l}_{2}}...,{{l}_{s}}l_{1}^{*}l_{2}^{*}....,l_{m}^{*}] \\ 
& =\left\{ \begin{aligned}
& 0,s\ne m \\ 
& \sum\limits_{\pi }{E[{{l}_{\pi (1)}}l_{1}^{*}]E[{{l}_{\pi (2)}}l_{2}^{*}],.....,E[{{l}_{\pi (s)}}l_{m}^{*}]},s=m \\ 
\end{aligned} \right.\\ 
\end{aligned}
\end{equation}
where $\left\{ {{l}_{i}}, i=1,2,.....s,.....,m \right\}$ are complex Gaussian random variables and  $\pi $ is apermutation of the set of integers $\left\{ 1,2,.....,s,....,m \right\}$\cite{magh31}. Due to Gaussian distribution of $y(t)$, (\ref{equ53}) is used to calculate ${{\phi }_{{{i}_{1}},{{i}_{2}}}}(t,\tau )=E[y{{(t)}^{{{i}_{1}}+1}}y{{(t-\tau )}^{{{i}_{2}}}}{{({{y}^{*}}(t))}^{{{i}_{1}}}}{{({{y}^{*}}(t-\tau ))}^{{{i}_{2}}+1}}]$ as
\begin{equation}\label{equ54}
{{\phi }_{{{i}_{1}},{{i}_{2}}}}(t,\tau )=\sum\limits_{\pi }{E[{{y}_{\pi (1)}}{{y}^{*}}(t)],...,E[{{y}_{\pi ({{i}_{1}})}}{{y}^{*}}(t)]}E[{{y}_{\pi ({{i}_{1}}+1)}}{{y}^{*}}(t-\tau )],...,E[{{y}_{\pi ({{i}_{1}}+{{i}_{2}}+1)}}{{y}^{*}}(t-\tau )]
\end{equation}
where $s=m={{i}_{1}}+{{i}_{2}}+1$ and ${{y}_{i}}=\left\{ \begin{aligned}
& y(t)\quad\quad\quad i=1,...,{{i}_{1}}+1 \\ 
& y(t-\tau )\quad i={{i}_{1}}+2,...,{{i}_{1}}+{{i}_{2}}+1 \\ 
\end{aligned} \right.$

By doing some manual calculation on (\ref{equ54}), ${{\phi }_{{{i}_{1}},{{i}_{2}}}}(t,\tau )$ is derived as 
\begin{equation}\label{equ55}
\begin{aligned}
{{\phi }_{{{i}_{1}},{{i}_{2}}}}(t,\tau )=&\sum\limits_{p=0}^{\min ({{i}_{1}},{{i}_{2}})}{\left( \begin{aligned}
	& {{i}_{2}}+1 \\ 
	& p+1 \\ 
	\end{aligned} \right)\left( \begin{aligned}
	& {{i}_{1}}+1 \\ 
	& p+1 \\ 
	\end{aligned} \right)\left( \begin{aligned}
	& {{i}_{2}} \\ 
	& p \\ 
	\end{aligned} \right)\left( \begin{aligned}
	& {{i}_{1}} \\ 
	& p \\ 
	\end{aligned} \right)(p+1)!(p)!({{i}_{2}}-p)!({{i}_{1}}-p)!} \\ 
& {{({{R}_{yy}}(t,\tau ))}^{p+1}}{{({{R}_{yy}}^{*}(t,\tau ))}^{p}}{{({{R}_{yy}}(t,0))}^{{{i}_{1}}+{{i}_{2}}-2p}} \\ 
\end{aligned}
\end{equation}

	\section{}
By considering (\ref{equ11}), (\ref{equ21}) and (\ref{equ51}), ${{\phi }_{{{i}_{1}},{{i}_{2}}}}(\tau )$ is derived. In the following, by taking FT of it, ${{\phi }_{{{i}_{1}},{{i}_{2}}}}(f)=\int\limits_{-\infty }^{\infty }{{{\phi }_{{{i}_{1}},{{i}_{2}}}}(\tau ){{e}^{-j2\pi f\tau }}}d\tau$  is calculated as 

       \begin{equation}\label{equ58}
     \begin{aligned}
      {{\phi }_{{{i}_{1}},{{i}_{2}}}}(f)=&\sum\limits_{p=0}^{\min ({{i}_{1}},{{i}_{2}})}{{{T}_{{{i}_{1}},{{i}_{2}}}}} \\ 
     & \times \left( \begin{aligned}
     & \sum\limits_{\upsilon =-\infty }^{\infty }{\sum\limits_{{{m}_{1}},...,{{m}_{p+1}}=0}^{M-1}\quad{\sum\limits_{m_{1}^{'},...,m_{p}^{'}=0}^{M-1}\quad{\sum\limits_{m_{1}^{''},...,m_{{{i}_{1}}+{{i}_{2}}-2p}^{''}=0}^{M-1}{{}}}}} \\ 
     & \left( \int\limits_{0}^{M{{T}_{s}}}{\sum\limits_{\upsilon =-\infty }^{\infty }{(\prod\limits_{j=1}^{p+1}{{{g}_{{{m}_{j}}}}(t-\upsilon {{T}_{B}})}\prod\limits_{{{j}^{'}}=1}^{p}{g_{m_{{{j}^{'}}}^{'}}^{*}(t-\upsilon {{T}_{B}})}{{\prod\limits_{{{j}^{''}}=1}^{{{i}_{1}}+{{i}_{2}}-2p}{\left| {{g}_{m_{{{j}^{''}}}^{''}}}(t-\upsilon {{T}_{B}}) \right|}}^{2}})}}dt \right) \\ 
     & \times \left( \int\limits_{-\infty }^{\infty }{\prod\limits_{j=1}^{p+1}{g_{{{m}_{j}}}^{*}(t-\tau -\upsilon {{T}_{B}})}\prod\limits_{{{j}^{'}}=1}^{p}{{{g}_{m_{{{j}^{'}}}^{'}}}(t-\tau -\upsilon {{T}_{B}})}{{e}^{-j2\pi f\tau }}}d\tau  \right) \\ 
     \end{aligned} \right) \\ 
     & \otimes \left( (\sum\limits_{{{k}_{1}},...,{{k}_{p+1}}=0}^{K-1}\quad{\sum\limits_{k_{1}^{'},...,k_{p}^{'}=0}^{K-1}{\delta (f-\dfrac{(\sum\limits_{j=1}^{p+1}{{{k}_{j}}}-\sum\limits_{{{j}^{'}}=1}^{p}{{{k}_{{{j}^{'}}}}-\dfrac{K-1}{2}})}{{{T}_{s}}})}} \right) \\ 
     \end{aligned}
       \end{equation} 
 where
\begin{equation}\label{equ76}
{{T }_{{{i}_{1}},{{i}_{2}}}}=\left( \begin{aligned}
& {{i}_{2}}+1 \\ 
& p+1 \\ 
\end{aligned} \right)\left( \begin{aligned}
& {{i}_{1}}+1 \\ 
& p+1 \\ 
\end{aligned} \right)\left( \begin{aligned}
& {{i}_{2}} \\ 
& p \\ 
\end{aligned} \right)\left( \begin{aligned}
& {{i}_{1}} \\ 
& p \\ 
\end{aligned} \right)(p+1)!(p)!({{i}_{2}}-p)!({{i}_{1}}-p)!{{(\alpha {{\overline{p}}_{x}})}^{{{i}_{1}}+{{i}_{2}}+1}}{{(K)}^{{{i}_{1}}+{{i}_{2}}-2p}}\frac{1}{M{{T}_{s}}}.
\end{equation}

 By defining ${{\tau }^{'}}=t-\tau -\upsilon {{T}_{B}}$  and using linear convolution formula, (\ref{equ58}) can be expressed as

  \begin{equation}\label{equ59}
\begin{aligned}
{{\phi }_{{{i}_{1}},{{i}_{2}}}}(f)=& \sum\limits_{p=0}^{\min ({{i}_{1}},{{i}_{2}})}{{{T}_{{{i}_{1}},{{i}_{2}}}}} \\ 
& \times \left[ \begin{aligned}
& \sum\limits_{{{m}_{1}},...,{{m}_{p+1}}=0}^{M-1}\quad{\sum\limits_{m_{1}^{'},...,m_{p}^{'}=0}^{M-1}\quad{\sum\limits_{m_{1}^{''},...,m_{{{i}_{1}}+{{i}_{2}}-2p}^{''}=0}^{M-1}{B(f)}}} \\ 
& \times \left( G_{_{{{m}_{1}}}}^{*}(f)\otimes ...\otimes G_{{{m}_{p+1}}}^{*}(f)\otimes {{G}_{m_{1}^{'}}}(-f)\otimes ...\otimes {{G}_{m_{p}^{'}}}(-f) \right) \\ 
\end{aligned} \right] \\ 
& \otimes \left( (\sum\limits_{{{k}_{1}},...,{{k}_{p+1}}=0}^{K-1}\quad{\sum\limits_{k_{1}^{'},...,k_{p}^{'}=0}^{K-1}{\delta (f-\dfrac{(\sum\limits_{j=1}^{p+1}{{{k}_{j}}}-\sum\limits_{{{j}^{'}}=1}^{p}{{{k}_{{{j}^{'}}}}-\dfrac{K-1}{2}})}{{{T}_{s}}})}} \right) \\ 
\end{aligned}
  \end{equation}
  where $B(f)$ is equal to

    \begin{equation}\label{equ60}
\begin{aligned}
 B(f)=&\int\limits_{0}^{M{{T}_{s}}}{\sum\limits_{\upsilon =-\infty }^{\infty }{(\prod\limits_{j=1}^{p+1}{{{g}_{{{m}_{j}}}}(t-\upsilon {{T}_{B}})}\prod\limits_{{{j}^{'}}=1}^{p}{g_{m_{{{j}^{'}}}^{'}}^{*}(t-\upsilon {{T}_{B}})}{{\prod\limits_{{{j}^{''}}=1}^{{{i}_{1}}+{{i}_{2}}-2p}{\left| {{g}_{m_{{{j}^{''}}}^{''}}}(t-\upsilon {{T}_{B}}) \right|}}^{2}}{{e}^{-j2\pi f(t-\upsilon {{T}_{B}})}})}}dt \\ 
& =\int\limits_{-\infty }^{\infty }{(\prod\limits_{j=1}^{p+1}{{{g}_{{{m}_{j}}}}(t)}\prod\limits_{{{j}^{'}}=1}^{p}{g_{m_{{{j}^{'}}}^{'}}^{*}(t)}{{\prod\limits_{{{j}^{''}}=1}^{{{i}_{1}}+{{i}_{2}}-2p}{\left| {{g}_{m_{{{j}^{''}}}^{''}}}(t) \right|}}^{2}}{{e}^{-j2\pi f(t)}})}dt \\ 
& =\left( \begin{aligned}
& {{G}_{{{m}_{1}}}}(f)\otimes...\otimes {{G}_{{{m}_{p+1}}}}(f)\otimes G_{m_{1}^{'}}^{*}(-f)\otimes ...\otimes G_{m_{p}^{'}}^{*}(-f)\otimes  \\ 
& G_{m_{1}^{''}}^{*}(-f)\otimes ...\otimes G_{m_{{{i}_{1}}+{{i}_{2}}-2p}^{''}}^{*}(-f)\otimes {{G}_{m_{1}^{''}}}(f)\otimes ...\otimes {{G}_{m_{{{i}_{1}}+{{i}_{2}}-2p}^{''}}}(f) \\ 
\end{aligned} \right). \\ 
\end{aligned}
    \end{equation} 
By considering  (\ref{equ59}) and (\ref{equ60}), ${{\phi }_{{{i}_{1}},{{i}_{2}}}}(f)$  is obtained.
\end{appendices}
	\bibliographystyle{IEEE}
	\bibliography{thesis-bib}

\end{document}